\documentclass[10pt]{blois07}
\usepackage{epsf,epsfig}
\usepackage{cite,./mcite}

\def\xb{\overline{x}}

\def\als{\alpha_s}
\def\vk{{\bf k}_{\perp}}

\def\gev{\,{\rm GeV}}

\begin{document}
\title{Blois07/EDS07 \\ Vector meson electroproduction  within GPD approach}
\author{S.V.\ Goloskokov}
\institute{Bogoliubov Laboratory of Theoretical Physics, Joint
Institute for Nuclear Research,\\ Dubna 141980, Moscow region,
Russia} \maketitle
\begin{abstract}
We analyze electroproduction  of light vector meson
 at small Bjorken $x$ within the generalized
parton distribution (GPD) approach. Calculation is based on the
modified perturbative approach, where the quark transverse degrees
of freedom in the hard subprocess are considered. Our results on
the cross section  are in fair agreement with experiment from
HERMES to HERA energies.
\end{abstract}

\section{Introduction}
\label{sec:xxx}

In this report, we investigate vector mesons electroproduction at
small Bjorken $x$ on the basis of the GPD approach
\cite{gk05,gk06}. At large $Q^2$ the leading order amplitude with
longitudinal photon and vector meson polarization (LL amplitude)
dominates and factorizes \cite{fact,*fact1} into a hard meson
leptoproduction off partons and GPDs,  Fig.1.
\begin{figure}[h]
\centering \mbox{\epsfysize=30mm\epsffile{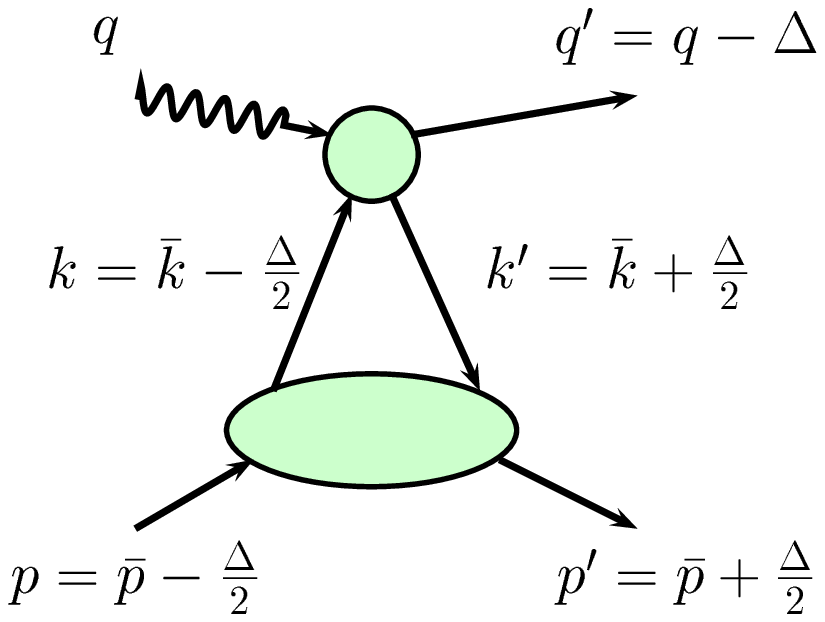}}
\caption{The handbag diagram for the meson electroproduction off
proton.} \label{kt_h}
\end{figure}
 Other transition amplitudes are suppressed by
powers of $1/Q$ and obey singularities \cite{mp,*anter} in the
collinear approximation. This leads to the problems with
factorization of these amplitudes.

In this report, we  analyse LL amplitude of vector meson
electroproduction at large $Q^2$. In contrast to \cite{gk05},
where the gluon predominant region $x \leq 0.01$ was considered,
we extend here our analysis to $x \leq 0.2$ \cite{gk06}. This
range covers the energy region from HERMES to HERA energies. Our
model is based on the modified perturbative approach (MPA)
\cite{sterman} which includes the quark transverse degrees of
freedom accompanied by Sudakov suppressions. The transverse quark
momentum regularizes the end-point region of the amplitudes. The
 $x \sim 0.2$ range study requires inclusion of the sea and
valence quark GPDs in our analysis. It is shown, that in our model
we obtain a fair description of  HERMES, H1 and ZEUS data
\cite{hermes, *hermes1, h1,*h1a,zeus,*zeusa} for electroproduced
$\rho$ and $\phi$ mesons at small $x$ \cite{gk06}.

\section{Leptoproduction of  Vector Mesons in the GPD approach}

The parton contribution to the photoproduction amplitudes
$\gamma^* p \to V p$ with positive proton helicity reads as a
convolution of the hard subprocess amplitude ${\cal H}^V$ and a
large distance unpolarized $ H^i$ and polarized $\widetilde{H}^i$
parton GPDs :
\begin{eqnarray}\label{amptt-nf-ji}
  {\cal M}^V_{\mu'+,\mu +} &=& \frac{e}{2}\, {\cal
  C}^{V}\, \sum_{\lambda}
         \int d\xb
        {\cal H}^{Vi}_{\mu'\lambda,\mu \lambda}
                                   H^i(\xb,\xi,t) ,
\end{eqnarray}
where  $i$ denotes the gluon and quark contribution,
 $\mu$ ($\mu'$) is the helicity of the photon (meson), $\xb$
 is the momentum fraction of the
parton with helicity $\lambda$, and the skewness $\xi$ is related
to Bjorken-$x$ by $\xi\simeq x/2$. The flavor factors are
$C_{\rho}=1/\sqrt{2}$ and ${ C}_{\phi}=-1/3$. The polarized GPDs
$\tilde H^i$ at small $x$ are much smaller than the unpolarized
GPDs $H^i$ and they are unimportant in the analysis of the cross
section.

 The subprocess amplitude ${\cal H}^V$ is represented as  the contraction of the hard
  part $F$ which is calculated perturbatively and the
non-perturbative meson  wave function $ \phi_V$ which depends on
the transverse quark momenta $k_\perp$ in the vector meson
\begin{equation}\label{hsaml}
  {\cal H}^V_{\mu'\lambda,\mu \lambda}=
\,\frac{2\pi \als(\mu_R)}
           {\sqrt{2N_c}} \,\int_0^1 d\tau\,\int \frac{d^{\,2} \vk}{16\pi^3}
            \phi_{V}(\tau,k^2_\perp)\;
                  F_{\mu^\prime\mu}^\pm .
\end{equation}
The wave function is chosen  in the simple Gaussian form
\begin{equation}\label{wave-l}
  \phi_V(\vk,\tau)\,=\, 8\pi^2\sqrt{2N_c}\, f_V a^2_V
       \, \exp{\left[-a^2_V\,
       \frac{\vk^{\,2}}{\tau\bar{\tau}}\right]}\, ,
\end{equation}
which leads after integration over  $k_\perp$ to the asymptotic
wave function. Here $\bar{\tau}=1-\tau$, $f_V$ is the decay
coupling constant and the $a_V$ parameter determines the value of
average transverse momentum of the quark.

In calculation of the subprocess   within the MPA \cite{sterman}
we keep the $k^2_\perp$ terms in the denominators of the hard
amplitudes.  The gluonic corrections in hard subprocess are
treated in the form of the Sudakov factors which additionally
suppress the end-point integration regions.

The GPD is a complicated function which depends on three
variables. We  use the double distribution form \cite{mus99}
\begin{equation}
  H_i(\xb,\xi,t) =  \int_{-1}
     ^{1}\, d\beta \int_{-1+|\beta|}
     ^{1-|\beta|}\, d\alpha \delta(\beta+ \xi \, \alpha - \xb)
\, f_i(\beta,\alpha,t),
\end{equation}
with the  distribution function
\begin{equation}
f_i(\beta,\alpha,t)= h_i(\beta,t)\,
                   \frac{\Gamma(2n_i+2)}{2^{2n_i+1}\,\Gamma^2(n_i+1)}
                   \,\frac{[(1-|\beta|)^2-\alpha^2]^{n_i}}
                          {(1-|\beta|)^{2n_i+1}}\,.
                          \end{equation}
Here
\begin{eqnarray}
& h_g(\beta,0)=|\beta|g(|\beta|)& n_g=2\nonumber\\
& h_{sea}^q(\beta,0)=q_{sea}(|\beta|) \mbox{sign}(\beta) & n_{sea}=2\nonumber\\
& h_{val}^q(\beta,0)=q_{val}(\beta) \Theta(\beta) & n_{val}=1,
\end{eqnarray}
where $g$ and $q$ are  ordinary gluon and quark PDFs.

For the  parton distribution the simple Regge ansatz is used
\begin{equation}\label{gd}
h_i(\beta,t)= e^{b_0 t}\beta^{-(\delta_i(Q^2)+\alpha'
t)}\,(1-\beta)^{2 n_i+1}\sum_{j=0}^3\,c_i^j\, \beta^{j/2}.
\end{equation}
The parameter $\delta_i(Q^2)$ is connected with the corresponding
Regge trajectory. For example, for gluon we have
\begin{equation}\label{delta}
\delta_g(Q^2) = \alpha_P(0)-1= 0.1 + 0.06 \ln{(Q^2/Q_0^2)},
\;Q_0^2= 4\mbox{GeV}^2,
\end{equation}
which determines the behavior of the gluon distribution at low $x$
and the energy dependence of the cross section at high energies.
The parameter $\alpha'$ in (\ref{gd}) is a slope of the Regge
trajectory $\alpha_i(t)=\alpha_i(0)+ \alpha'_i t$.  The other
parameters in (\ref{gd}) are taken from comparison with the CTEQ6M
parameterization ~\cite{CTEQ}.

The valence quark sea  differs from the strange sea. The simple
model is used
\begin{equation}
H^u_{sea}=H^d_{sea}=\kappa_s H^s_{sea},
\end{equation}
which is in correspondence with the CTEQ6 results. The flavor
symmetry breaking factor is chosen in the form
\begin{eqnarray}\label{kappas}
\kappa_s=1+0.68/(1+0.52 \ln{(Q^2/Q_0^2)}
\end{eqnarray}
which fits well  CTEQ6M PDFs.
\begin{figure}[h!]
\begin{center}
\begin{tabular}{ccc}
\mbox{\epsfig{figure=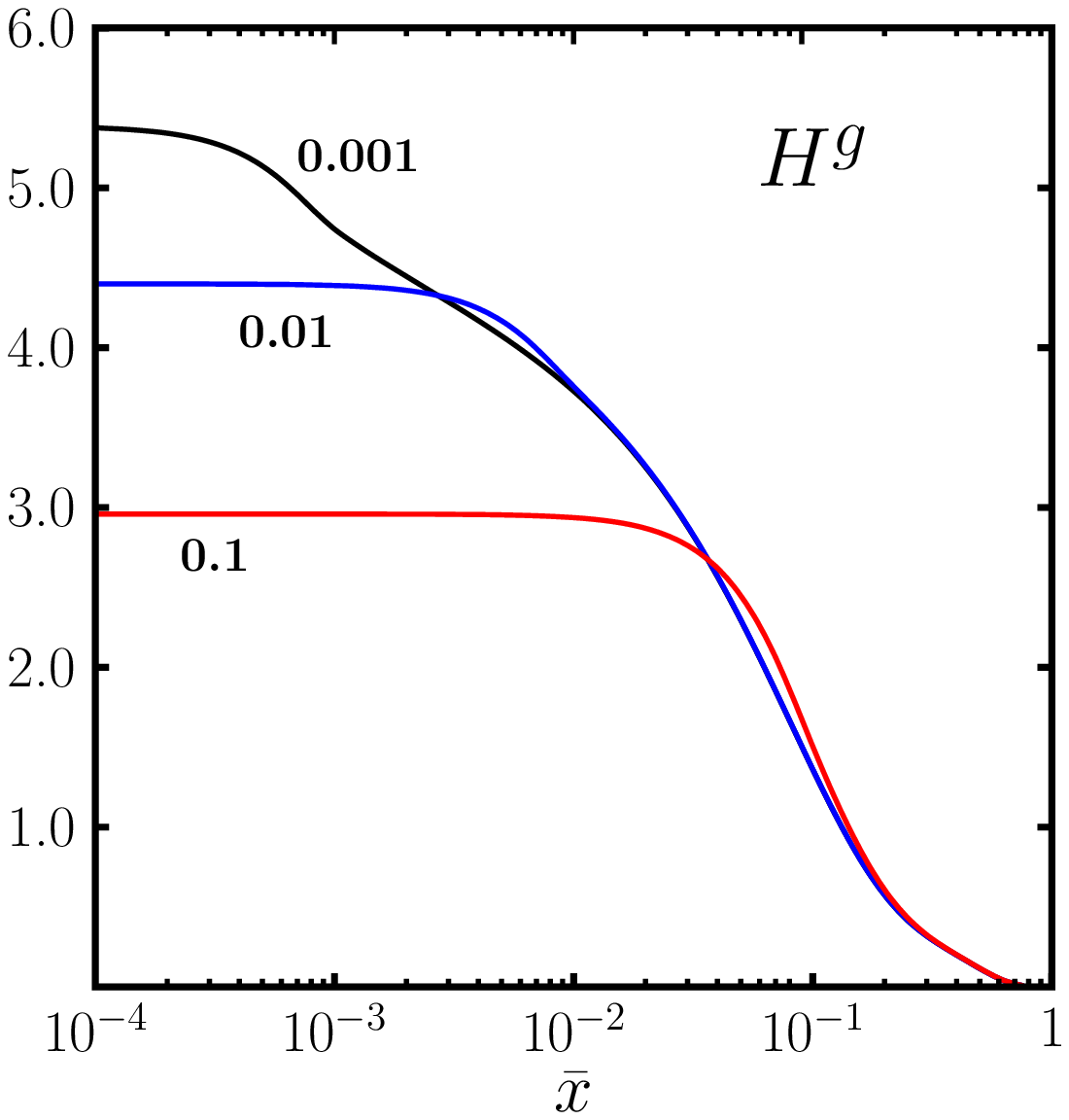,width=4.6cm,height=4.8cm}}&
\mbox{\epsfig{figure=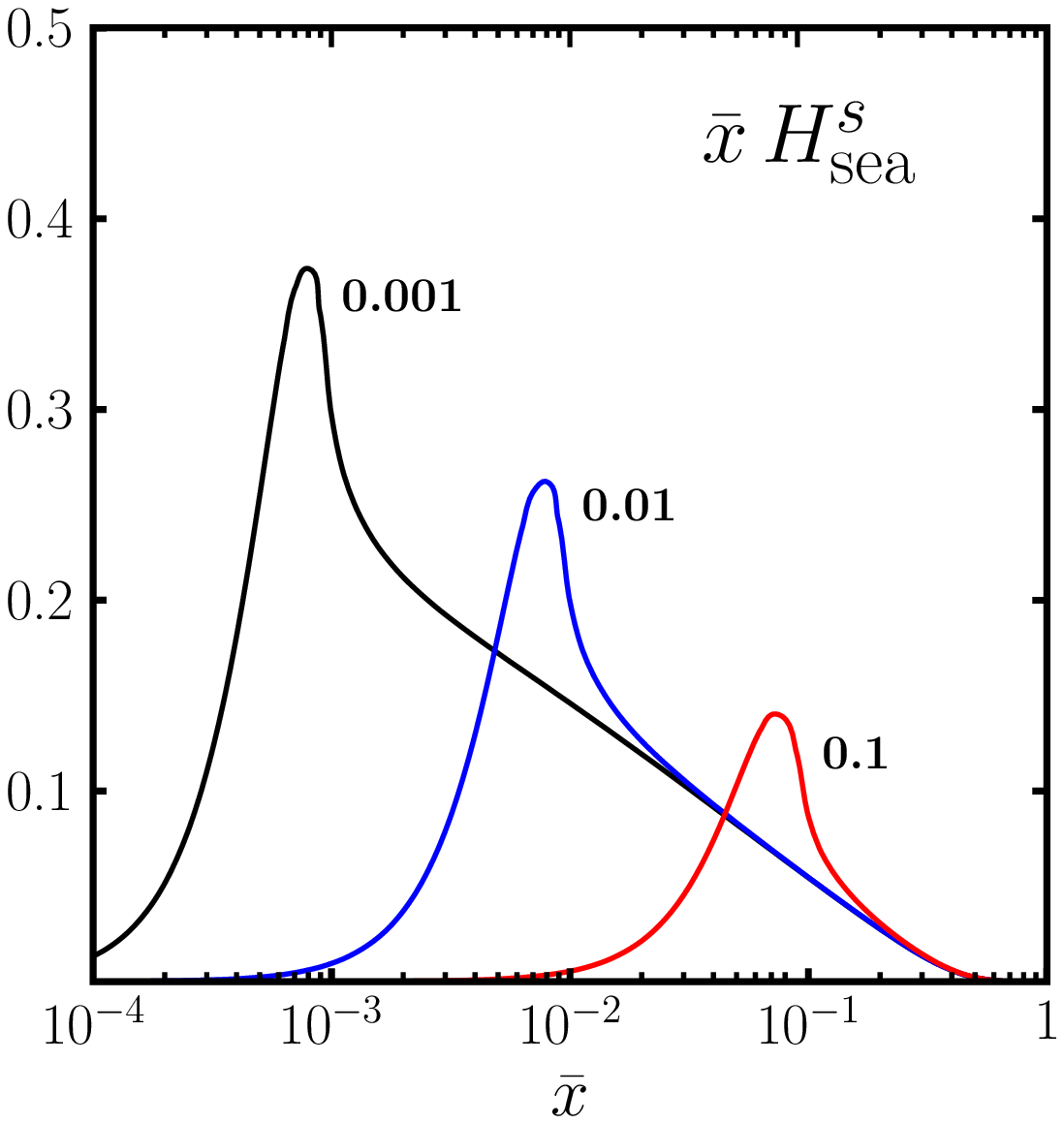,width=4.6cm,height=4.8cm}}&
\mbox{\epsfig{figure=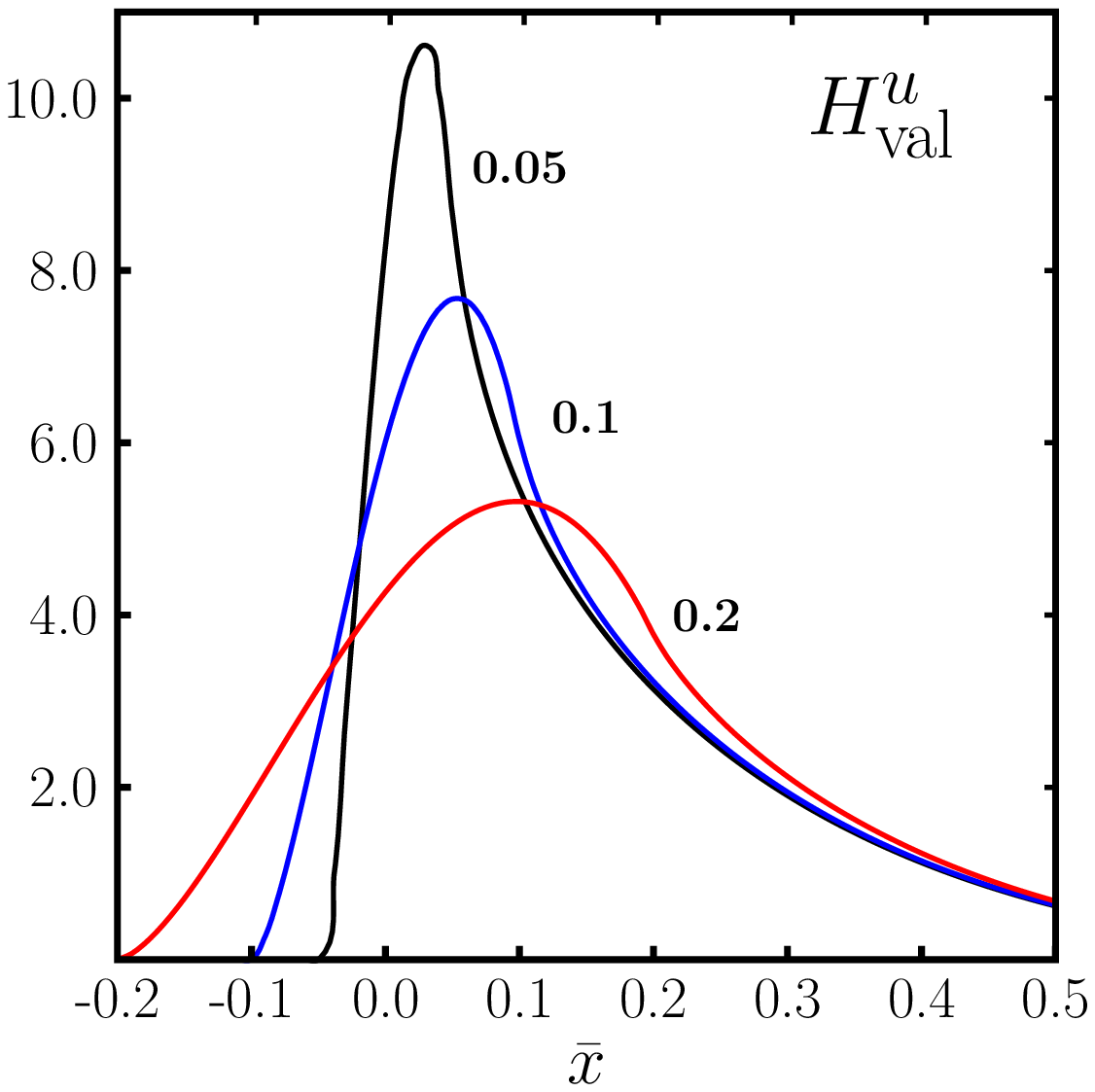,width=4.6cm,height=4.7cm}}\\
{\bf(a)}& {\bf(b)}& {\bf(c)}
\end{tabular}
\end{center}
\caption{ GPDs {\bf(a)} $H^g$, {\bf(b)} $x H^s_{sea}$, {\bf(c)}
$H^u_{val}$ via $x$ for some values of skewness. GPDs are shown at
$t =0$ and scale $Q^2=4\mbox{GeV}^2$ }
\end{figure}

The model results for gluon and quark GPDs for different values of
skewness are shown in Fig. 2.

\section{Longitudinal cross section}
Now we have all ingredients to calculate cross sections.
Estimations for the vector meson production
 are obtained using $f_{\rho L}=0.209\gev$, $a_{\rho
L}=0.75\gev^{-1}$; $f_{\phi L}=0.221\gev$; $a_{\phi
L}=0.7\gev^{-1}$. The value of the diffractive peak slope can be
found in \cite{gk06}. The longitudinal cross section for the
$\rho$ and $\phi$ production integrated over $t$  is shown in Fig.
3 at HERA energies. In this energy range the valence quark effects
are unimportant. In Fig. 3a, the cross section of $\rho$
production is shown together with the individual contributions to
the cross section: gluon contribution,  the gluon-sea-quark
interference, and quark contribution. It can be seen that a
typical contribution of the interference to $\sigma_L$ does not
exceed 50\% with respect to the gluon one. Thus, the gluon term
really gives the predominant contribution to the cross section
\cite{diehl} and we find  good agreement of our results with the
H1 and ZEUS experiments \cite{h1,zeus} at HERA.

The model results for the  $\phi$ production cross section shown
in Fig. 3b  are consistent with the H1 and ZEUS data
\cite{h1,zeus}. In $\phi$ production the gluon-sea quark
interference contribution to the cross section does not exceed
25\%. Note that the uncertainties in the  GPDs provide errors in
the cross section about $25-35 \%$ which are shown in Fig. 3b for
$\phi$ production. The  $\rho$ production cross section in Fig. 3a
has the similar uncertainties. They are of the same order of
magnitude as the gluon-sea interference. The leading twist results
which do not consider effects of transverse quark motion, are
presented in Fig. 3b too. One can see that the $k_\perp^2/Q^2$
corrections in the hard amplitude denominators are extremely
important at low $Q^2$. They decrease the cross section by a
factor of about 10 at $Q^2 \sim 3\mbox{GeV}^2$. The role of these
corrections at $Q^2 \sim 40 \mbox{GeV}^2$ is not so
essential-about 40\%.

\begin{figure}[h!]
\begin{center}
\begin{tabular}{cc}
\mbox{\epsfig{figure=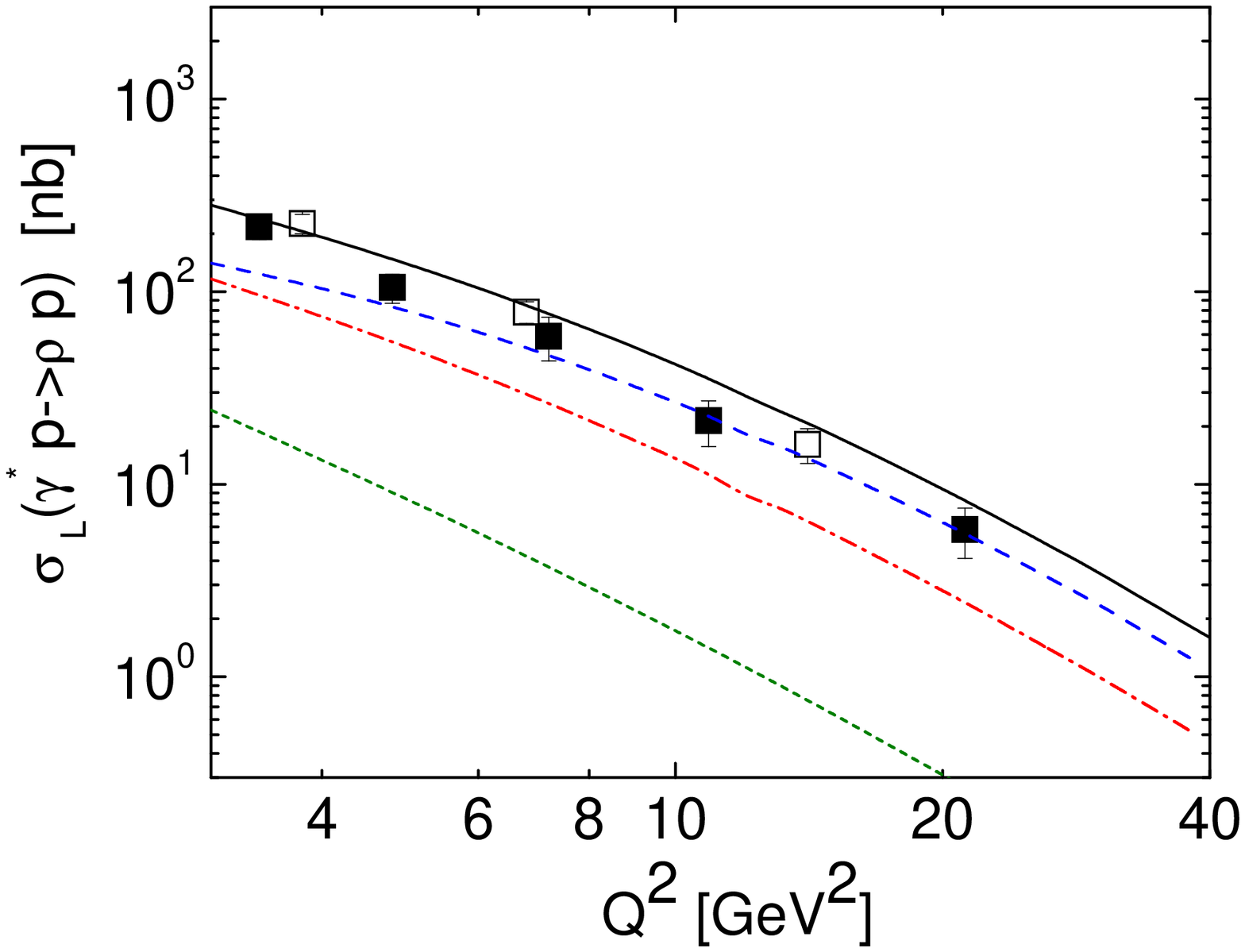,width=7.2cm,height=6.2cm}}&
\mbox{\epsfig{figure=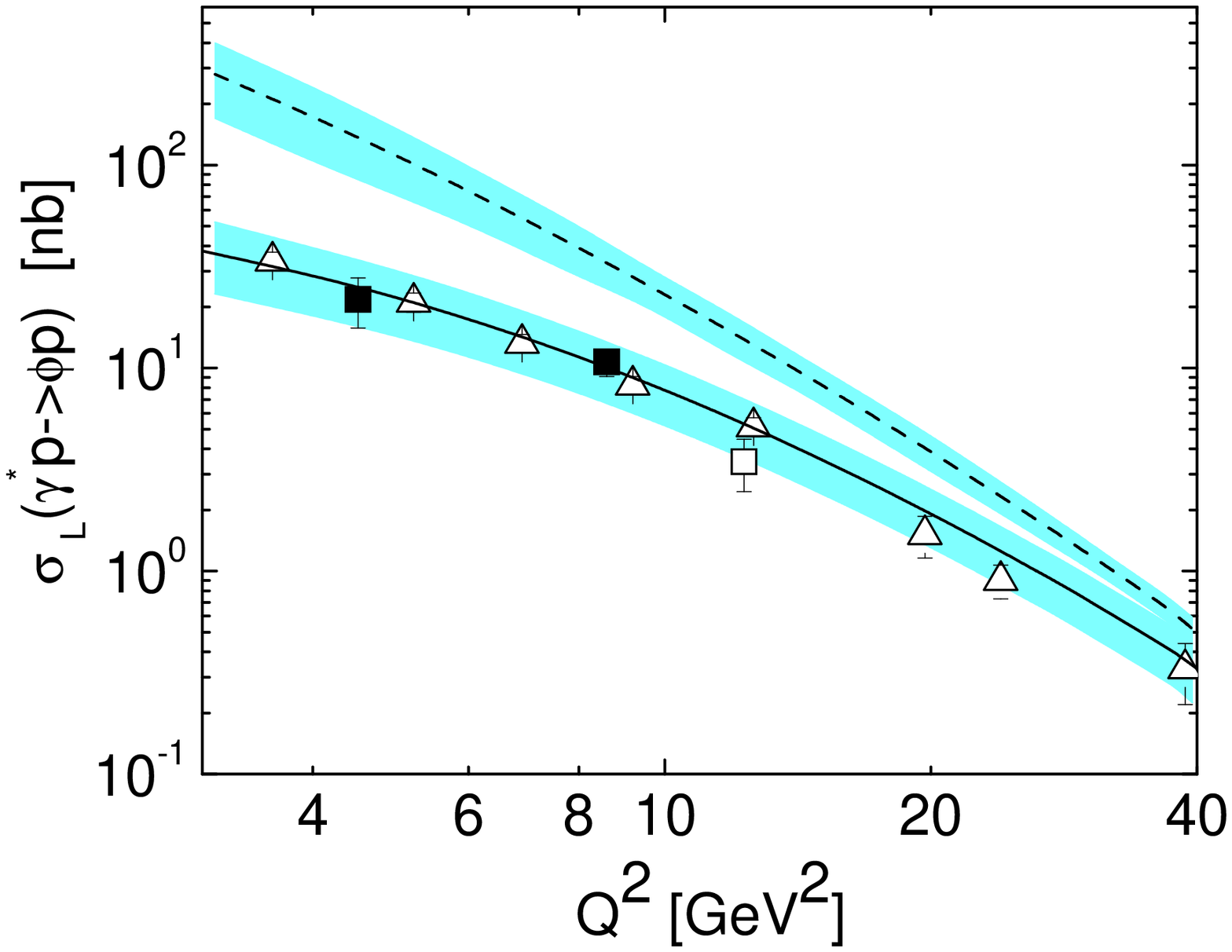,width=7.2cm,height=6.2cm}}\\
{\bf(a)}& {\bf(b)}
\end{tabular}
\end{center}
\caption{ {\bf(a)} Longitudinal cross sections of $\rho$
production at $W=75 \mbox{GeV}$. Full line cross section, dashed-
gluon contribution, dashed-dot - gluon-sea interference, dotted
line -sea contribution. {\bf(b)} Full line- longitudinal cross
sections of $\phi$ production at $W=75 \mbox{GeV}$ with error band
from CTEQ PDF uncertainties. Dashed line -leading twist results.
Data are from H1 and ZEUS. }
\end{figure}

Let us discuss the energy dependence of the cross section. At
small $x$ where only gluon and sea contribute they behave as
\begin{equation}\label{sigma}
\sigma_L \propto W^{4 \delta(Q^2)},
\end{equation}
where the power $\delta$   is determined in (\ref{delta}). At
larger $x$ the valence quark contribution should play an important
role. In Fig. 4 a, we show our results for the $\rho$- production
cross section at $Q^2=4 \mbox{GeV}^2$ in a wide energy range.
Together with the gluon contribution, the gluon + sea and
interference of valence quark with gluon + sea plus valence quark
contribution to the cross section are shown. It can be seen that
for  energies above $W \geq 10\mbox{GeV}$ the gluon and sea
effects well reproduce the cross section.
\begin{figure}[h!]
\begin{center}
\begin{tabular}{cc}
\mbox{\epsfig{figure=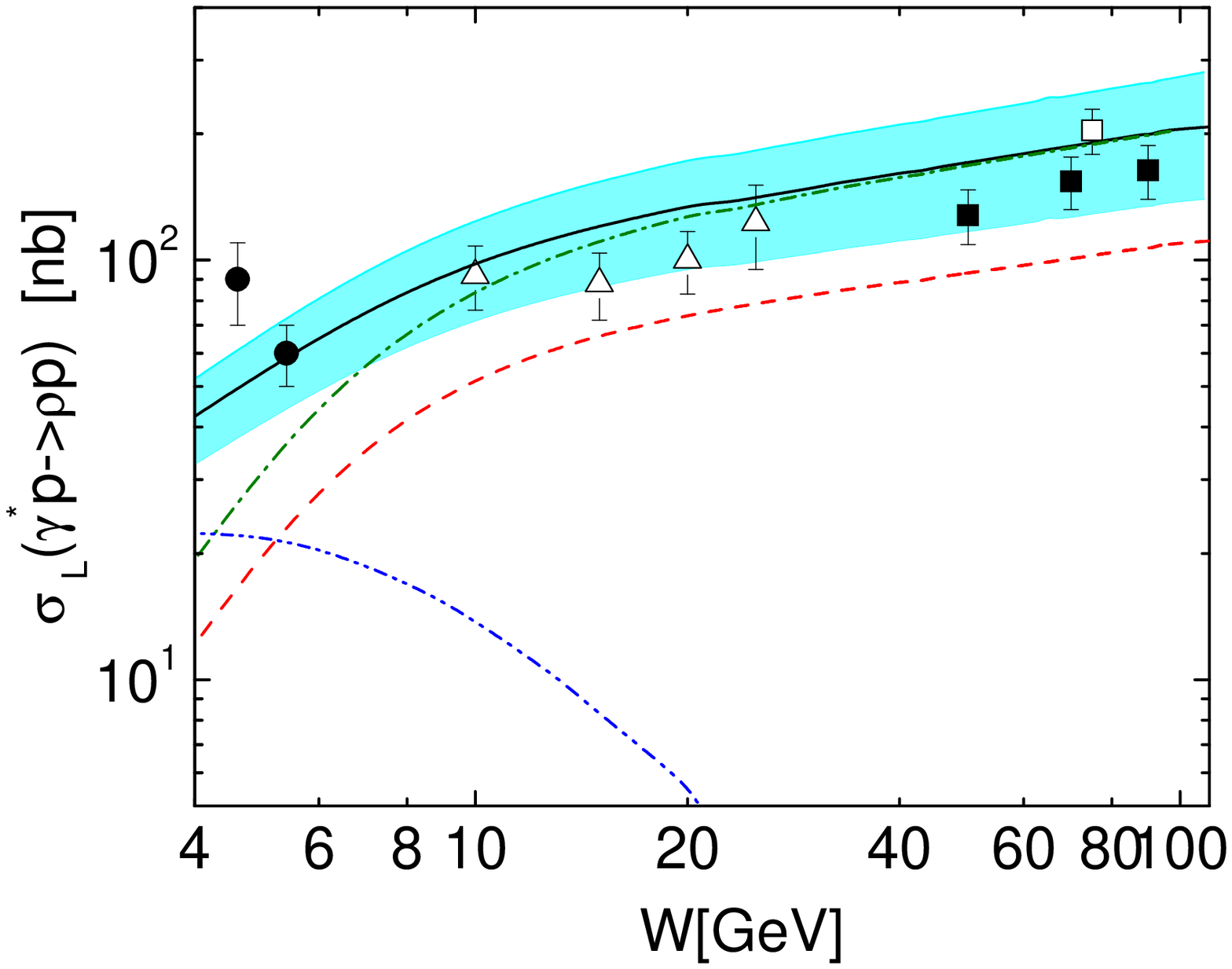,width=7.2cm,height=6.2cm}}&
\mbox{\epsfig{figure=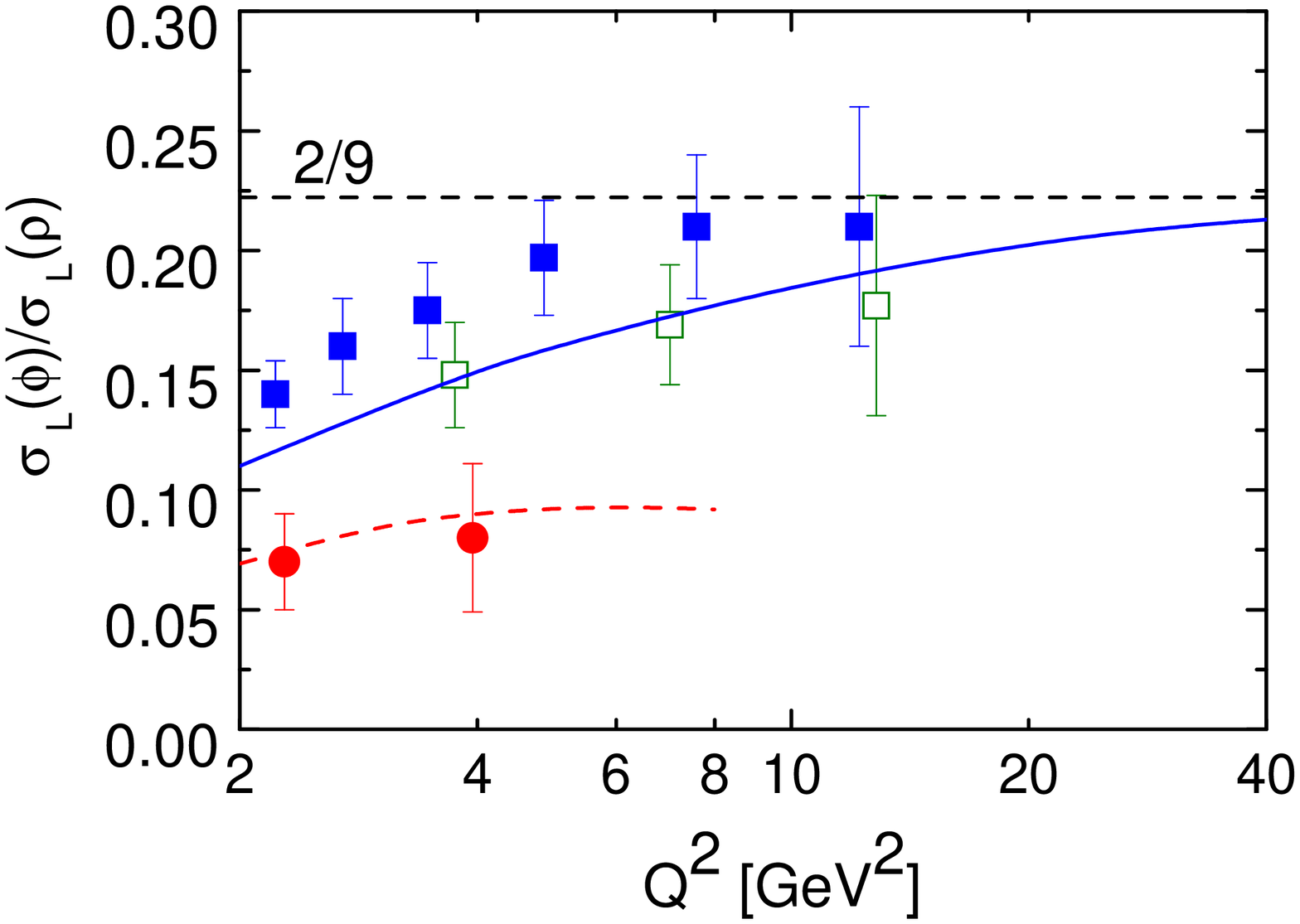,width=7.2cm,height=6.3cm}}\\
{\bf(a)}& {\bf(b)}
\end{tabular}
\end{center}
\caption{ {\bf(a)} Longitudinal cross sections of $\rho$
production at $Q^2=4 \mbox{GeV}^2$ as a function of $W$. Full line
cross section, dashed- gluon contribution, dashed-dot - gluon+sea,
dashed dot dotted- (gluon+sea)-valence interference plus valence
contribution. {\bf(b)} The ratio of $\phi/\rho$ cross sections via
$Q^2$. Full line- $W=75 \mbox{GeV}$, dashed line -$W=5
\mbox{GeV}$. Data are from H1 and ZEUS and HERMES.}
\end{figure}
At HERMES energies $W \sim 5\mbox{GeV}$ the valence quarks
contribution is important. For $\rho$ production the interference
of valence quarks with gluons and sea contribution give  of about
40\% contribution to the cross section at  $W \sim 5\mbox{GeV}$.
At COMPASS energies
 $W \sim 10\mbox{GeV}$ the valence quarks give
 only about 10\% contribution to the cross
section. Thus, COMPASS physics is very  close to asymptotic HERA
energies.

The ratio of the $\phi/\rho$ cross sections at HERA energies $W
=75 \mbox{GeV}$ is shown in Fig. 4.b. It is obvious that if the
valence quark sea does not contribute (or sea is symmetric), this
ratio is determined by the flavor factors in (\ref{amptt-nf-ji})
and should be equal to $\sigma(\phi)/\sigma(\rho)=2/9$. The HERA
data show a strong deviation of this ratio from $2/9$ value. In
our model, this violation   at HERA energies and low $Q^2$ finds a
natural explanation  by the flavor symmetry breaking factor
(\ref{kappas}) effect. At high $Q^2$ the $\kappa_s$ factor  goes
to 1 and the ratio of the cross section is close to the
$\sigma(\phi)/\sigma(\rho)=2/9$ limit. Thus, we can conclude that
the $Q^2$ dependence of the $\sigma(\phi)/\sigma(\rho)$ ratio is
completely determined by the flavor symmetry breaking factor
$\kappa_s$. It cannot be explained  if one does not consider the
quark sea contribution.  At HERMES energies the valence quarks
contribution in $\rho$ production  gives an additional suppression
the of $\sigma(\phi)/\sigma(\rho)$ ratio -see Fig. 4b.

 \section{Conclusion or Summary}
We have analyzed electroproduction of light  mesons at small
Bjorken-$x$ in the handbag model  where the process amplitudes are
factorized into the GPDs and a partonic subprocess. The subprocess
was calculated \cite{gk06} within the modified perturbative
approach where the transverse momenta of the quark and antiquark
as well as Sudakov corrections were taken into account. These
effects suppress the contributions from the end-point regions
where one of the partons entering into the meson wave function
becomes soft and factorization breaks down in the collinear
approximation. It is found that the GPD approach gives a fine
description of the longitudinal cross section  for light meson
production. The power corrections $\sim k_\perp^2/Q^2$ in
propagators of the hard amplitude play an extremely important role
at low $Q^2$. Inclusion of these corrections gives a possibility
to describe experimental data properly.

The gluonic contribution plays an essential role for all energies
$W>5 \mbox{GeV}$ in vector meson electroproduction. The gluon-sea
interference is about 30(50)\% for $\phi$ ($\rho$) production.
Valence quarks contribute only for $W< 10 \mbox{GeV}$. For the
$\rho$ production at HERMES energies $W\sim 5 \mbox{GeV}$, valence
quarks give about 40\% effect in the cross section. At COMPASS
$W\sim 10 \mbox{GeV}$ their contribution is about 10\% only. The
flavor symmetry breaking of the sea naturally explain the
deviation of the  $\sigma(\phi)/\sigma(\rho)$ ratio  from the
asymptotic  limit equal to 2/9 at HERA energies at low $Q^2$.

 Thus,
we can conclude that in different energy ranges, information about
quark and gluon GPDs can be extracted from the cross section of
the vector meson electroproduction. This reaction   at small $x$
and large $Q^2$ is an excellent tool to study the gluon and quark
GPDs.

 \bigskip

 This work is supported  in part by the Russian Foundation for
Basic Research, Grant 06-02-16215  and by the Heisenberg-Landau
program.

\end{document}